# Macroscopic phase segregation in superconducting $K_{0.73}Fe_{1.67}Se_2$ as seen by muon spin rotation and infrared spectroscopy

C. N. Wang,[1] P. Marsik,[1] R. Schuster,[1] A. Dubroka,[1,2] M. Rössle,[1] Ch. Niedermayer,[3] G. D. Varma,[4] A. F. Wang,[5] X. H. Chen,[5] T. Wolf,[6] and C. Bernhard[1]

[1] *University of Fribourg, Department of Physics and Fribourg Centre for Nanomaterials, Chemin du Musée 3, CH-1700 Fribourg, Switzerland*

[2] *Department of Condensed Matter Physics, Faculty of Science, Masaryk University, and Central European Institute of Technology, Kotlářská 2, CZ-61137 Brno, Czech Republic*

[3] *Laboratory for Neutron Scattering, Paul Scherrer Institut, CH-5232 Villigen, Switzerland*

[4] *Department of Physics, Indian Institute of Technology Roorkee, Roorkee 247667, India*

[5] *Hefei National Laboratory for Physical Sciences at Microscale and Department of Physics, University of Science and Technology of China, Hefei, Anhui 230026, China*

[6] *Karlsruhe Institut für Technologie, Institut für Festkörperphysik, D-76021 Karlsruhe, Germany*

**Abstract**

Using muon spin rotation ($\mu$SR) and infrared spectroscopy we investigated the recently discovered superconductor $K_{0.73}Fe_{1.67}Se_2$ with $T_c \approx 32$ K. We show that the combined data can be consistently described in terms of a macroscopically phase segregated state with a matrix of ~88% volume fraction that is insulating and strongly magnetic and inclusions with a ~12% volume fraction which are metallic, superconducting and non-magnetic. The electronic properties of the latter, in terms of the normal state plasma frequency and the superconducting condensate density, appear to be similar as in other iron selenide or arsenide superconductors.

The recent discovery of superconductivity with $T_c$ above 30 K in the alkali metal intercalated iron-chalcogenide with the nominal composition $K_{0.8}Fe_{2-y}Se_2$[1,2] has led to great efforts to better understand their unusual electronic and magnetic properties.[3] Superconductivity is reported to occur here in the presence of a very strong antiferromagnetic order with a Néel temperature of $T_N \approx 560$ K and a large magnetic moment of ~3.6 $\mu_B$ per Fe ion.[4] Meanwhile it is well established that high temperature superconductivity (HTSC) in the cuprates[5-7] and iron arsenides[8-11] occurs in close proximity to an antiferromagnetic state. Nevertheless, the static magnetic order is usually strongly suppressed or even entirely absent in the superconducting part of the phase diagram. Another unusual feature concerns the extremely low electronic conductivity of these iron selenide superconductors.[12,13] It implies that the concentration of itinerant charge carriers is more than an order of magnitude smaller than in their cuprate and iron arsenide counterparts, where it is already considered to be very low. This has led to speculations that the mechanism of superconductivity in these chalcogenides may be different and even more unconventional than in the cuprate and iron arsenide high temperature superconductors. Alternatively, there exists mounting evidence that this material may be spatially inhomogeneous consisting of a matrix that is strongly antiferromagnetic and insulating and inclusions that are metallic and superconducting.[14-18] Since a coherent superconducting response is observed in electric transport and macroscopic magnetization measurements,[2] the latter phase must either amount to a significant volume fraction or else its inclusions must be specifically shaped and arranged such that percolation is achieved along certain pathways.

In the following we present a combined muon-spin-rotation ($\mu$SR) and infrared (IR) spectroscopy study which supports a macroscopic phase segregation scenario. In good agreement with recent reports,[14,17,18] our $\mu$SR data establish the presence of two phases that are either strongly magnetic or entirely non-magnetic and superconducting with volume fractions of ~88% and ~12%, respectively. They yield an estimate of the magnetic penetration depth of $\lambda_{ab} \approx 270$ nm that is similar as in the other iron arsenide, selenide or cuprate superconductors with corresponding $T_c$ values.[12,13,19,20] We also show that by adopting the volume fractions as obtained from $\mu$SR, a reasonable description of the infrared spectra can be obtained with an effective medium approximation (EMA). The model uses an insulating matrix and metallic inclusions for which the Drude response has a similar plasma frequency as in other iron selenide or arsenide superconductors. The shape of the metallic inclusions appears to be elongated such that percolation is achieved despite the low volume fraction.

Superconducting single crystals with a composition of $K_{0.73}Fe_{1.67}Se_2$ (KFS_SC) were grown in Hefei, China as described in Ref. 2, 21. Corresponding non-superconducting and insulating crystals with a nominal composition $K_{0.75}Fe_{1.60}Se_2$ (KFS_I) were prepared in Karlsruhe, Germany. They were grown from K (3N5), Fe (3N), and Se (6N) at a ratio of 0.75:1.6:2 in a vertical Bridgman setup. The starting mixture was filled into a tipped $Al_2O_3$ crucible that was sealed in a steel container before the crystal growth was carried out in a tubular furnace by cooling from 1050 °C to 770 °C at a rate of 0.4 °C/h. Electric transport measurements confirmed the weakly metallic and superconducting properties of KFS_SC with a superconducting transition at $T_c \approx 32$ K as shown in Fig. 1.

The $\mu$SR measurements were performed on crystals with an area of about 5 mm$^2$ using the GPS setup at the $\pi$M3 beamline of the Paul Scherrer Institut (PSI) in Villigen, Switzerland, which provides a beam of 100% spin polarized muons. $\mu$SR measures the time evolution of the spin polarization, $P(t)$, of the implanted muon ensemble via the time-resolved asymmetry, $A(t)$, of the muon decay positrons.[22] The technique is well suited to studies of magnetic and superconducting materials as it allows a microscopic determination of the internal field distribution and can give direct access to the volume fractions of the superconducting and magnetic phases. The positive muons are implanted into the bulk of the sample and stop at well-defined interstitial lattice sites.[23] The muon ensemble is distributed in a layer of 100-200 $\mu$m thickness and therefore probes a representative part of the sample volume. Each muon spin precesses in the local magnetic field $B_\mu$ with a precession frequency of $v_\mu = \gamma_\mu \cdot B_\mu/2\pi$, where $\gamma_\mu = 2\pi \cdot 135.5$ MHz/$T$ is the gyromagnetic ratio of the muon.

The IR spectroscopy was performed on freshly cleaved surfaces of the same crystals that were used for the $\mu$SR study. Details of the infrared ellipsometry and reflection techniques and the analysis of the combined date are described in Refs. 24-26. Special care was taken to avoid any degradation of the surface due to the contact with the ambient. The samples were mounted in argon gas atmosphere using a glove box and were quickly inserted into the ellipsometry cryostat which then was immediately evacuated. For the IR reflectivity measurements the crystals were even in-situ cleaved inside the cryostat at a temperature of about 5 K. Corresponding reflectivity measurements on crystals that were ex-situ cleaved before mounting them in the cryostat confirmed that the weak electronic response of the superconducting crystals as shown in the following does not originate from a degraded surface but is instead an intrinsic property of the bulk of these samples.

Figure 2 summarizes our transverse-field (TF) $\mu$SR experiments on samples KFS_SC and KFS_I at an external magnetic field $H_{ext} = 0.1$ kOe that was applied parallel to the $c$-axis of the crystals. Figure 2(a) shows the muon spin polarization, $P(t)/P(t = 0)$, for KFS_SC at 38 K just above $T_c = 32$ K. The major part of the signal exhibits an extremely rapid depolarization that is not even captured within the experimental resolution of 0.625 ns. The remaining part is oscillatory and relaxes only very slowly. The fast relaxing component was previously reported for superconducting crystals with a nominal composition of $Cs_{0.8}Fe_2Se_2$[27] and interpreted in terms of the antiferromagnetic order of the large Fe moments. The observation of two components with such drastically different relaxation rates is suggestive of a spatially inhomogeneous state. Based on the amplitude of the $\mu$SR signals we estimate that the strongly magnetic phase involves ~88% of the sample volume while only a minor fraction of ~12% remains non-magnetic and is most likely superconducting. This estimate takes into account that about 2-3% of the slowly depolarizing $\mu$SR signal arises from muons that stop outside the sample in the sample holder or the cryostat walls. Recently a very similar result has been reported in Ref. 28. Our interpretation is confirmed by the corresponding data on the non-superconducting sample KFS_I as shown in Fig. 2(c). Here only the 2-3% of the signal due to the background muons remains slowly relaxing, whereas the entire signal from the muons stopping in the sample is now rapidly depolarized.

The alternative explanation of the TF-$\mu$SR data in terms of two different muon stopping sites in the unit cell seems rather unlikely. It would require a second muon site where the local magnetic field vanishes despite the very large Fe moments. Furthermore, since the non-magnetic signal occurs only for the superconducting crystals, the presence of this highly symmetric muon site would have to be linked to the appearance of a metallic and superconducting state.

Our interpretation that the slowly relaxing component arises from a certain fraction of the sample volume that is non-magnetic and superconducting is supported by the temperature dependence of the muon spin relaxation rate. As shown in Figs. 2(b) and 2(d) it exhibits a pronounced increase below $T_c$ = 32 K that is characteristic of the formation of a superconducting vortex lattice. As shown by the solid lines in Fig. 2, the TF-$\mu$SR data were analyzed with the following function:

$$P(t) = P(0) \cdot \left[ A_f \cos(\gamma_\mu B_{\mu,f} t + \varphi) \exp(-\lambda_f t) + A_s \cos(\gamma_\mu B_{\mu,s} t + \varphi) \exp\left(-\frac{\sigma_s^2 t^2}{2}\right) \right]$$

eq. (1)

Here $A_f$ and $A_s$ are the amplitudes of the fast and slowly relaxing components, respectively, $\gamma_\mu$ the gyromagnetic ratio of the muon, $B_\mu$ the local magnetic field at the muon site, $\lambda_f$ the exponential relaxation rate of the fast component, $\varphi$ the initial phase of the muon spins, and $\sigma_s$ the Gaussian relaxation rates of the slowly relaxing component. In the normal state the small and $T$-independent value of $\sigma_s$ is determined by the nuclear magnetic moments. The order parameter like increase of $\sigma_s$ below $T_c$ signifies the formation of a superconducting vortex lattice. Assuming that the size of the superconducting inclusions is large enough to enable the formation of a regular vortex lattice in their interior, we derive from the low temperature value of $\sigma_s \approx 1.35$ $\mu s^{-1}$ an estimate of the in-plane magnetic penetration depth of $\lambda_{ab} \approx 270$ nm. Notably this value of $\lambda_{ab}$ agrees well with the one reported in Ref. 28 and it is also similar as in other iron arsenide[11,29-31] and cuprate superconductors with a comparable $T_c$ value.[19,32]

The superconducting origin of the enhanced relaxation below $T_c$ has been furthermore established with a so-called pinning experiment which reveals the presence of a strongly pinned superconducting vortex lattice. Figure 3 displays so-called $\mu$SR lineshapes that were obtained from a fast Fourier transformation of the TF-$\mu$SR time spectra. The open symbols show the $\mu$SR lineshape as measured after the sample was cooled to 1.6 K in an applied magnetic field of $H_{appl} = 750$ Oe. It has the expected characteristic shape with a very narrow peak at $H_{appl}$ that arises from the background muons that stop outside the sample and a broader main peak that is shifted to lower fields (diamagnetic shift). The somewhat asymmetric shape with a tail towards higher fields originates from the muons that stop near to the vortex cores. The second lineshape as shown by the solid symbols has been obtained after the applied magnetic field was reduced by 100 Oe from 750 to 650 Oe while the temperature was kept at 1.6 K. While the narrow peak due to the background muons follows this reduction of $H_{appl}$, the broader main part of the $\mu$SR lineshape remains virtually unchanged. This characteristic behavior which highlights that the magnetic flux density in the sample remains unchanged is the hallmark of a type-II superconductor with a strongly pinned vortex lattice. It clearly demonstrates that the non-magnetic regions of KFS_SC become superconducting below $T_c$ = 32 K. The observation of a well-developed and strongly pinned vortex lattice is also consistent with the assumption that these superconducting regions are fairly sizeable, i.e. larger than the magnetic penetration depth $\lambda_{ab}$. Finally we note that the TF-$\mu$SR data do not provide any specific information about the properties of the strongly magnetic fraction. In particular, due to the extremely large relaxation rate, we cannot tell whether or not this fraction is superconducting.

In the following we present the infrared spectroscopy data which provide important, complementary information about the electronic properties. In particular, we show that they establish that the matrix of KFS_SC is insulating and non-superconducting while the inclusions are metallic with a plasma frequency that is similar as in other iron selenides or arsenides where superconductivity is a bulk phenomenon. Figure 4(a) displays the $T$-dependence of the in-plane reflectivity, $R_{ab}$, for KFS_SC in the far-infrared range. As shown by the dotted lines, a Hagen-Rubens relation has been used to extrapolate the data toward zero frequency. The corresponding real part of the optical conductivity, $\sigma_{1,ab}$, as obtained from a Kramers-Kronig analysis is shown in Fig. 4(b). Shown by the full symbols are the values of the dc conductivity, $\sigma_{dc}$, from the resistivity data in Fig. 1. The corresponding values of $\sigma_{1,ab}(\omega\rightarrow 0)$ as obtained from the low-frequency extrapolation of our optical data (dotted lines) are consistently somewhat higher. Nevertheless, given the various uncertainties of these transport and optical measurements, the agreement is reasonable and it validates our low-frequency extrapolation procedure.

In good agreement with previous reports,[13,17,33] we find that these infrared spectra contain only weak signatures of a metallic response. The value of the plasma frequency of the free carriers of $\omega_{pl} \approx$ 100-150 cm$^{-1}$, as estimated from the position of the reflection edge in $R_{ab}$ or from an analysis of the optical conductivity with a Drude-Lorentz model, is indeed very small. For a spatially homogeneous sample it would amount to a free carrier concentration that is about two orders of magnitude smaller than in any of the iron selenide/telluride, iron arsenide or cuprate superconductors.[25,30,34,35] It would also be incompatible with the magnetic penetration depth of $\lambda_{ab}$ = 270 nm or a plasma frequency of the superconducting condensate of $\omega_{pl}^{SC}$ = 5894 cm$^{-1}$ as obtained from the $\mu$SR data. In addition to the weakly conducting response, the IR spectra exhibit a broad band centered at about 300 cm$^{-1}$ which is most pronounced at low temperature. Since this band is much broader than the phonon lineshapes, it has most likely an electronic origin.

In the following we show that this inconsistently weak electronic response and the band at 300 cm$^{-1}$ can be naturally accounted for if one considers that the electronic state may be spatially inhomogeneous. Figure 5 shows that the normal state optical spectra can be very well described in terms of an effective medium approximation (EMA) model. In modeling the spectra, we assumed that the system is composed of an insulating matrix and metallic inclusions with volume fractions of 88% and 12%, respectively, as deduced from the $\mu$SR data. The EMA model assumes that the inclusions are randomly oriented ellipsoids with an aspect ratio, $Q$, that is treated as a fitting parameter.[36] The dielectric function of the insulating matrix was determined from the optical measurements on sample KFS_I as shown in Figs. 4(d)-(f). The dielectric function of the conducting inclusions was modeled with two Drude terms, a narrow one to account for the coherent response and a broad one to represent the less coherent background. The fitted plasma frequencies of these Drude-peaks, $\omega_{pl,Dn}$ and $\omega_{pl,Db}$, are 3873 cm$^{-1}$ and 3162 cm$^{-1}$, respectively. In addition we introduced a broad Lorentzian oscillator with a width of 1000 cm$^{-1}$ that is centered at 5000 cm$^{-1}$.

We found that in order to reproduce both the very weak free carrier response and the 300 cm$^{-1}$ band, the response has to be modeled as a volume average of two different EMA-models, the so-called Maxwell-Garnett (MG) model and the Bruggeman (B) effective medium approximation. The necessity of the combination arises from the well known properties of the two approaches, i.e., that the MG-EMA describes very well the resonant plasmonic oscillations occurring in isolated inclusions which in our model give rise to the 300 cm$^{-1}$ band, whereas the low frequency percolative behavior is reasonably well described by the B-EMA. Specifically, the simulated spectra shown in Fig. 5 were obtained by assuming that percolation is achieved in ~90% of the sample volume (described by the B-model) while in ~10% the metallic inclusions are disconnected by the insulating matrix (accounted for by the MG-model). Note that in both the B- and the MG-EMA models we used the same volume fraction of the metallic inclusions of 12% as obtained from the $\mu$SR experiments. The aspect ratio $Q$ was found to be 0.085 from both MG-EMA and B-EMA models suggesting that the metallic inclusions have a very elongated, needle-like shape. Notably, a similar shape of the inclusions was obtained with an electron backscattering analysis in Ref. 18. In reality there is likely a variation in the concentration and/or the shape of these metallic inclusions that determines whether percolation is achieved in certain parts of the sample. Nevertheless, we did not allow for such a variation since, as shown below, the presented model describes reasonably well on a qualitative and even quantitative level the key features of the infrared response. Concerning the phonon resonances we note that these have not been modeled but are entirely determined by the measured spectra of the insulating sample KFS_I.

Figure 5 shows that our EMA model enables a reasonable fitting of the normal state spectra. Concerning the electronic response, it accounts well for the small value of the apparent plasma frequency and it reproduces the broad electronic band around 300 cm$^{-1}$ that becomes pronounced below 100 K. Figure 5 furthermore shows that the B-MG-EMA model accounts for the $T$-dependence in the normal state. The width of the broad Drude peak, $\Gamma_{Db}$, as well as the unscreened plasma frequencies of the narrow and broad Drude-peaks, $\omega_{pl,Dn}$ and $\omega_{pl,Db}$, were fixed at 300, 3873 and 3162 cm$^{-1}$, respectively. The resulting total, unscreened plasma frequency of the two Drude components of $\omega_{pl,D} \approx 5000$ cm$^{-1}$ is of the same order of magnitude as the one reported for e.g. bulk iron selenide and arsenide superconductors.[25,37-41] The only parameter that is strongly varied is the width of the narrow Drude-peak, $\Gamma_{Dn}$, whose $T$-dependence is shown in Fig. 6. The very small value of $\Gamma_{Dn} \approx 5$ cm$^{-1}$ at 35 K just above $T_c$ is a remarkable feature. While a significantly larger value of about 90 cm$^{-1}$ has been reported for BaFe$_{1-x}$Co$_x$As$_2$,[25] a similarly small value of the scattering rate at low temperature and low frequency was obtained for FeTe$_{0.55}$Se$_{0.45}$ with a so-called extended Drude model analysis.[41] Figure 6 also compares the dc conductivity, $\sigma_{dc}$, as obtained from the resistivity data in Fig. 1 with the low-frequency extrapolated value of $\sigma_{1,ab}(\omega \rightarrow 0)$ that is predicted by our EMA model. The values agree reasonably well, even at $T = 35$ K where the fit yielded an unusually small scattering rate.

Concerning the infrared-active phonon modes, our model describes the modes at 104, 150, 210 and 240 cm$^{-1}$ whereas it does not account for the ones at 120 and 280 cm$^{-1}$. This indicates that the metallic inclusions may have a different structure than the insulating matrix. Evidence for an insulating matrix of K$_x$Fe$_4$Se$_5$ with a $\sqrt{5} \times \sqrt{5}$ Fe-vacancy ordering and superconducting inclusions of a K$_x$Fe$_2$Se$_2$ phase containing stoichiometric FeSe layers, has indeed been reported from scanning tunneling spectroscopy measurements[42,43] on samples that are from the same growth batch as KFS_SC. Alternatively, the additional phonons may arise from a minor structural difference between the insulating matrix of the KFS_SC sample and the KFS_I sample that was used for the modeling.

Finally, we discuss the changes of the infrared spectra in the superconducting state. As shown in Fig. 4(a), the reflectivity at 6 K as compared to the one at 35 K exhibits a weak yet noticeable increase below ~100 cm$^{-1}$ which may well be the signature of a superconducting energy gap. The real part of the optical conductivity at low frequency is suppressed here at the cost of a zero-frequency delta function that accounts for the loss-free response of the superconducting condensate. The latter also gives rise to an enhancement of the low-frequency inductive response which shows up as a decrease of the real part of the dielectric function towards negative values. Nevertheless an unambiguous identification of such a superconducting condensation effect is complicated by the circumstance that a $T$-dependent increase of the low-frequency reflectivity occurs already in the normal state. As was demonstrated above, it is related to the narrowing of the Drude-response. Therefore it is difficult to ascertain which part of the observed changes below $T_c$ is caused by superconductivity. In addition, one has to keep in mind that for this inhomogeneous system, the effective response of the metallic/superconducting phase is strongly modified with respect to the one of a bulk system. For example, a significant part of the spectral weight of the Drude-response as well as of the delta-function due to the superconducting condensate does not show up at the origin but instead becomes part of the band at 300 cm$^{-1}$. Based on this EMA modeling, it is therefore rather difficult to determine the finer details of the superconductivity-induced changes of the optical response. Nevertheless, as shown in the following it can still be used to discuss some exemplary cases.

As a first case (Model-A) we assumed that the broad Drude-peak remains unaffected by the superconducting transition, whereas all the charge carriers involved in the narrow Drude-peak condense and give rise to a delta function at the origin of the conductivity spectrum. The corresponding conductivities at 35 K just above $T_c$ and well below $T_c$ are shown in Figs. 7(a) and 7(b), respectively. The shaded area in the latter indicates the so-called "missing spectral weight" that is transferred to the delta-function at zero frequency and forms the superconducting condensate. The comparison between the fit and the experimental data is shown in Figs. 8(a)-(c). The obtained values of the superconducting plasma frequency and the magnetic penetration depth are $\omega_{pl,SC} \approx 3873$ cm$^{-1}$ and $\lambda_{ab} \approx 410$ nm, respectively. The former value is about 2.3 times smaller than the one obtained above from the $\mu$SR data. Nevertheless, in comparing these values we remark that the condensate density as deduced from $\mu$SR experiments is frequently found to be quite a bit larger than the one derived from infrared spectroscopy.[44]

As a second case (Model-B), we assumed that both the broad and the narrow Drude components develop an isotropic superconducting gap. The gap magnitude we assumed to be $2\Delta \approx 16$ meV $\approx 130$ cm$^{-1}$ as reported from recent angle-resolved photo-emission spectroscopy (ARPES) measurements.[4,45-47] The corresponding conductivities at 35 K just above $T_c$ and at 6 K well below $T_c$ are displayed in Fig. 7(c). Shown by the shaded area is the missing spectral weight that is transferred to the delta-function at zero frequency. The comparison between the fit and the experimental data is shown in Figs. 8(d)-(f). The isotropic superconducting gap introduces a sharp edge in the reflectivity around 50 cm$^{-1}$ that is not observed in the experimental spectra where the reflectivity rises more gradually toward low frequency and remains below unity in the entire measured range of $\omega > 30$ cm$^{-1}$. This implies that the low-frequency optical response does not become fully gapped in the superconducting state, i.e. the optical conductivity remains finite as shown in Fig. 8(e). Based on the present data we cannot answer the question whether this is due to superconducting gap that is anisotropic in momentum space, a very small or even vanishing gap on one of the conduction bands, or simply some inhomogeneity in the properties of the superconducting inclusions. We note that Model B yields estimates of $\omega_{pl,SC} \approx 4160$ cm$^{-1}$ and $\lambda_{ab} \approx 380$ nm that are somewhat closer to the $\mu$SR values. However, it needs to be remarked that Model B is likely to overestimate the superconducting condensation density. Finally we note that the sharp reflectivity edge in the modeled spectra occurs at a significantly lower frequency of 50 cm$^{-1}$ than the gap at $2\Delta \approx 130$ cm$^{-1}$ where the reflectivity edge would locate for the case of a homogeneous, bulk superconductor. This downward shift of the reflection edge is a consequence of the B-EMA approach, it would not occur if we only used the MG-EMA model.

In summary, using muon spin rotation ($\mu$SR) and infrared spectroscopy we investigated the magnetic and electronic properties of K$_{0.73}$Fe$_{1.67}$Se$_2$ single crystals with $T_c = 32$ K and compared them to the ones of non-superconducting crystals with a nominal composition of K$_{0.75}$Fe$_{1.6}$Se$_2$. The combined data provided evidence that the crystals are spatially inhomogeneous with a majority phase (matrix) that is insulating and strongly magnetic and embedded inclusions that are non-magnetic, metallic and superconducting. The $\mu$SR data established that the latter amount to about 12% of the sample volume. The analysis of the infrared data with a model based on the effective medium approximation revealed that the metallic inclusions have an elongated, almost needle-like shape and that percolation between them is achieved in most (but not all) of the sample volume. It also showed that the plasma frequency of the free carriers as well as of the superconducting condensate density are sizeable, e.g. on the same order of magnitude as in other iron selenide and arsenide superconductors.

**Acknowledgement**

The work was partially performed at the infrared beamline of the ANKA synchrotron source at KIT Karlsruhe, Germany and at the Swiss muon Source (S$\mu$S) at the Paul Scherrer Institut, Villigen, Switzerland. The work at University of Fribourg has been supported by the Swiss National Science Foundation (SNF) grant 200020-129484, by the NCCR MaNEP and by the project no. 122935 of the Indo-Swiss Joint Research Program. The work at the Central European Institute of Technology was supported by the project no. CZ.1.05/1.1.00/02.0068. The work in Hefei was supported by the Natural Science Foundation of China, the Ministry of Science and Technology of China (973 project no. 2006CB60100) and the Chinese Academy of Sciences. The work in Germany has been supported by the Deutsche Forschungsgemeinschaft through SPP 1458. R. Schuster appreciates funding through the DFG via SCHU/2584/1-1. The financial supports from D.S.T. (Government of India) and M.H.R.D. (Government of India) are highly acknowledged.

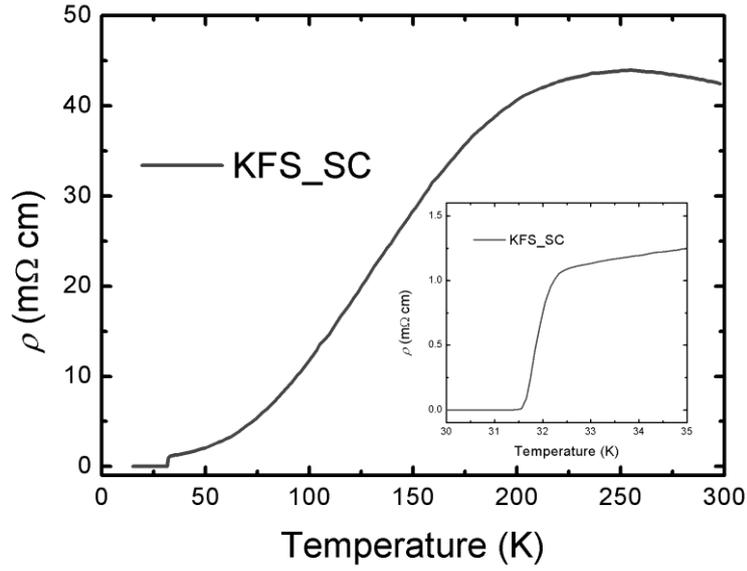

**Figure 1:** Temperature dependence of the in-plane resistivity of a KFS_SC single crystal from the same growth batch as the samples used for the $\mu$SR and infrared measurements.

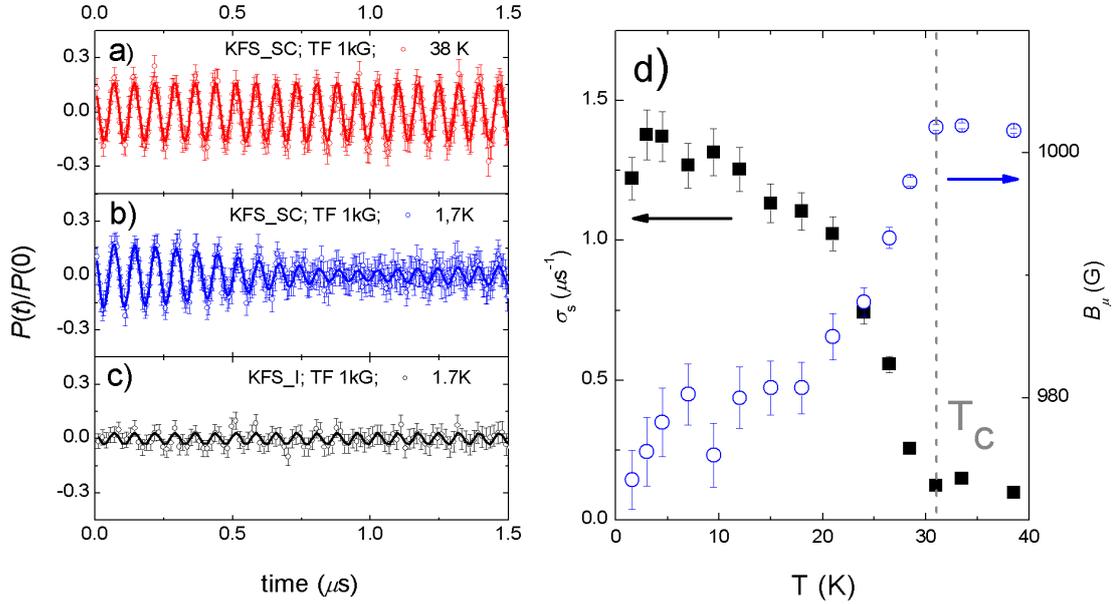

**Figure 2:** Representative TF-$\mu$SR data (symbols) of **(a)** and **(b)** KFS_SC just above and below $T_c \approx 32$ K, respectively, and **(c)** of KFS_I. Solid lines show the fits obtained with the function in eq. (1). **(d)** Temperature dependence of the local field, $B_\mu$, and the relaxation rate, $\sigma_s$, of the slowly relaxing signal of sample KFS_SC.

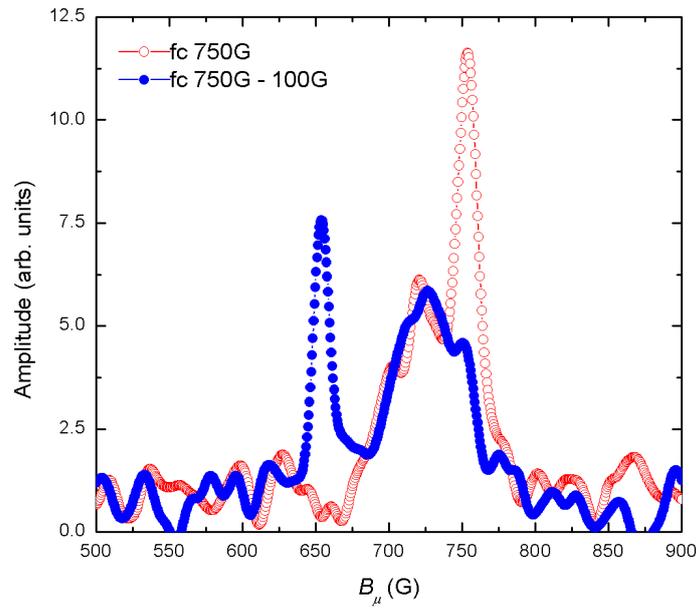

**Figure 3:** TF-$\mu$SR line shapes showing the distribution of local magnetic fields during a so-called pinning experiment as described in the text.

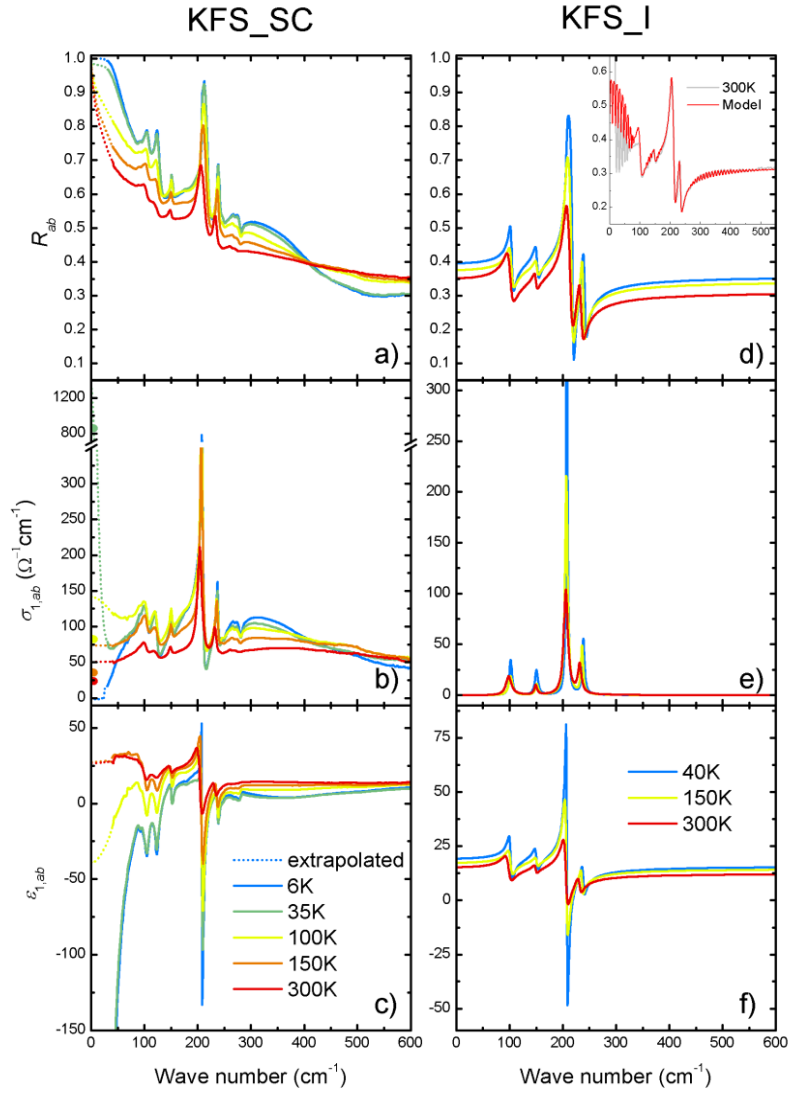

**Figure 4:** Temperature dependence of the far-infrared spectra of KFS_SC in terms of **(a)** the in-plane polarized reflectivity, $R_{ab}$, **(b)** the real part of the optical conductivity, $\sigma_{1,ab}$, and **(c)** the dielectric function, $\varepsilon_{1,ab}$. Shown by the symbols in (a) is the dc conductivity, $\sigma_{dc}$, deduced from the resistivity data in Fig. 1(a). **(d)**-**(f)** Corresponding spectra for KFS_I. Since the crystal was insulating and thus transparent, the spectra were corrected for the interference effects that arise from multiple reflections as shown in the inset of (d).

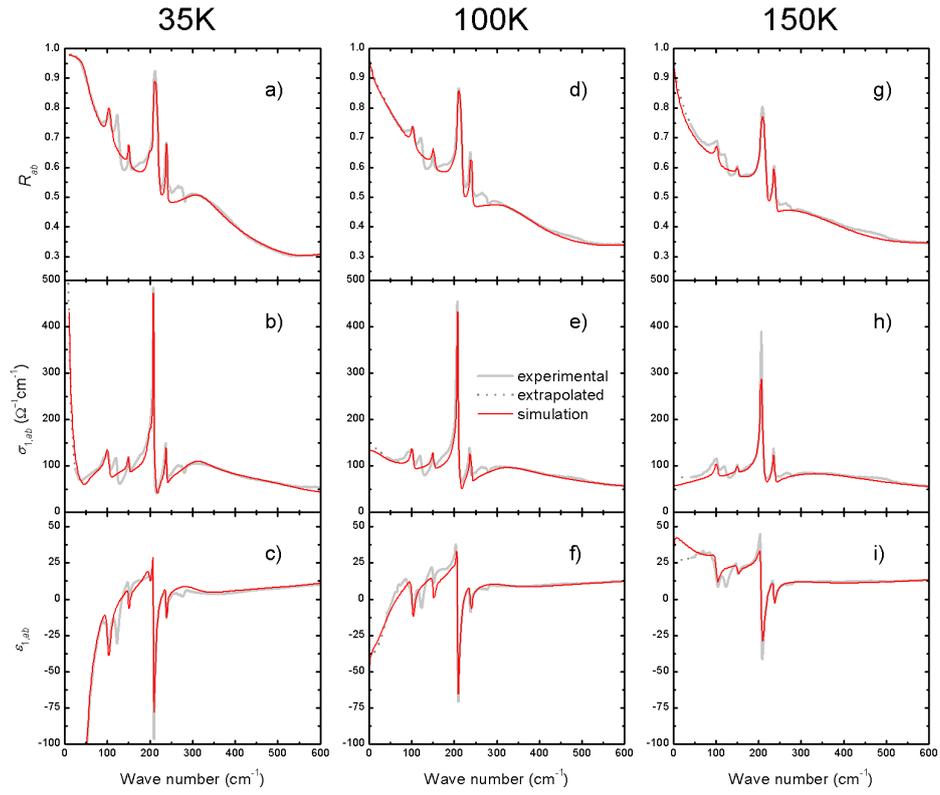

**Figure 5:** Comparison of the experimental data in the normal state of KFS_SC with the fits based on the effective medium approximation (EMA) model as described in the text.

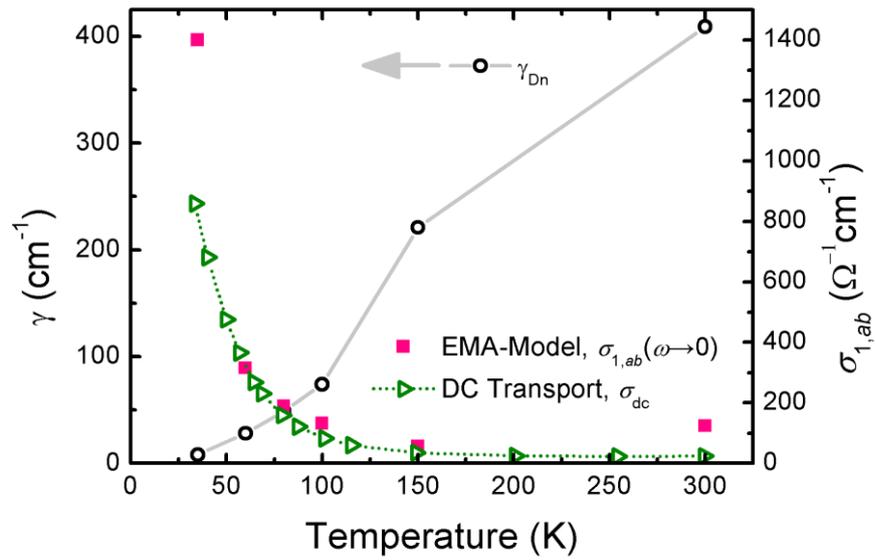

**Figure 6:** Temperature dependence of the scattering rate of the narrow Drude-peak, $\Gamma_{Dn} = 1/\tau_{Dn}$, as obtained from the fitting the with EMA model that is described in the text. Also shown for comparison is the dc conductivity, $\sigma_{dc}$, obtained from the resistivity data in Fig. 1(a) and the extrapolated value of $\sigma_1(\omega\to 0)$ obtained with the EMA model.

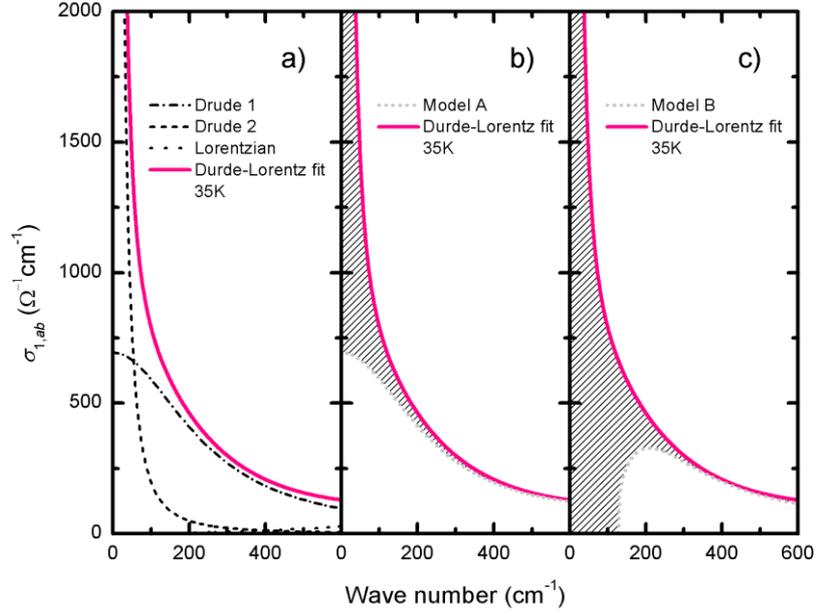

**Figure 7: (a)** Sketch of the two Drude- and the Lorentzian components that were used to fit the data at 35 K just above $T_c$. **(b)** and **(c)** sketch of the changes due to the onset of superconductivity that were assumed in the context of Model A and Model B, respectively. The shaded area indicates the missing spectral weight that is redistriubed towards the origin where it accounts for the response of the superconducting condensate.

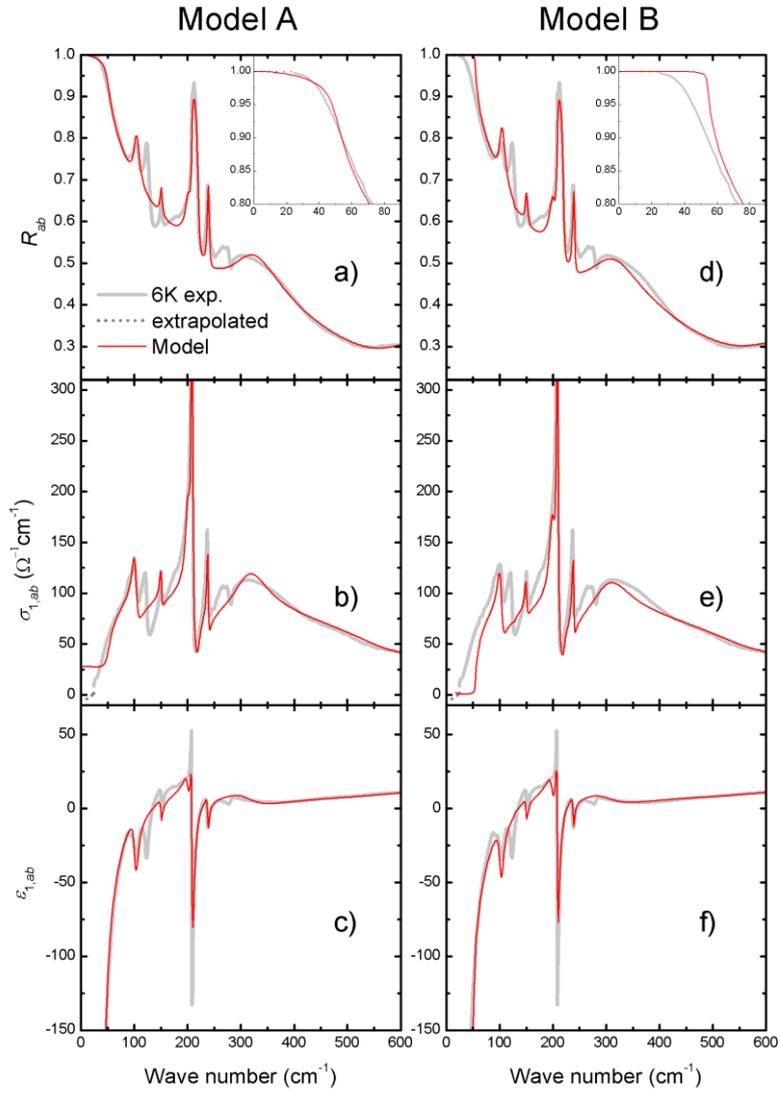

**Figure 8:** Comparison of the experimental data at 6 K $\ll T_c \approx$ 32 K of KFS_SC with the EMA fits using Model A and Model B, respectively. The two models are outlined in the text and sketched in Figs. 7(b) and 7(c).